\documentclass[runningheads]{llncs}
\usepackage[T1]{fontenc}
\usepackage{hyperref}

\usepackage{graphicx}
\usepackage[para]{footmisc}

\newcommand{\quotes}[1]{``#1''}

\begin{document}
\title{ForestQB: An Adaptive Query Builder to Support Wildlife Research}
%
%\titlerunning{Abbreviated paper title}
% If the paper title is too long for the running head, you can set
% an abbreviated paper title here
%
\author{Omar Mussa\inst{1,3} \and
Omer Rana\inst{1} \and
Benoît Goossens\inst{2}\and
Pablo Orozco-terWengel\inst{2}\and
Charith Perera\inst{1}}

% \author{Omar Mussa\inst{1}\orcidID{0000-0001-8614-6550} \and
% Omer Rana\inst{1}\orcidID{0000-0003-3597-2646} \and
% Benoît Goossens\inst{1}\orcidID{0000-0003-2360-4643}\and
% Pablo Orozco-terWengel\inst{1}\orcidID{0000-0002-7951-4148}\and
% Charith Perera\inst{1}\orcidID{0000-0002-0190-3346}}

%
\authorrunning{O. Mussa et al.}
% First names are abbreviated in the running head.
% If there are more than two authors, 'et al.' is used.
%
\institute{School of Computer Science and Informatics, Cardiff University, United Kingdom \\
\email{\{mussao, ranaof, pererac\}@cardiff.ac.uk}
\and
School of Biosciences, Cardiff University, United Kingdom\\
\email{\{goossensbr, orozco-terwengelpa\}@cardiff.ac.uk}
\and
College of Computing and Informatics, Saudi Electronic University, Saudi Arabia\\
\email{o.mousa@seu.edu.sa}
}
\maketitle              % typeset the header of the contribution
\begin{abstract}
This paper presents ForestQB, a SPARQL query builder, to assist Bioscience and Wildlife Researchers in accessing Linked-Data. As they are unfamiliar with the Semantic Web and the data ontologies, ForestQB aims to empower them to benefit from using Linked-Data to extract valuable information without having to grasp the nature of the data and its underlying technologies. ForestQB is integrating Form-Based Query builders with Natural Language to simplify query construction to match the user requirements.

\keywords{Linked-Data  \and Visual Querying \and SPARQL \and Query Builders.\\
\textbf{Paper type:} Demo (available at \url{https://iotgarage.net/demo/forestQB})

}

\end{abstract}
\section{Introduction}
Publishing the data as a Linked-Data using Semantic Web technologies is beneficial for machine learning as well as information retrieval \cite{Malyshev2018}. While the data will be easily accessible by machines, Humans can also benefit from accessing the data by using a query language such as SPARQL, which is the recommended query language for querying RDF triplestore.

In the field of Bioscience and Wildlife conservation, researchers tend to collect data using various sensors such as temperature, location and speed. Therefore, hundreds of gigabytes were collected over the years that would be extremely valuable if stored as a knowledge graph in an RDF triplestore. However, users usually feel intimidated to use Linked-Data as they are obliged to understand SPARQL and the underlying data structure \cite{Warren2020}. In order to encourage these Bioscience researchers to adopt semantic web technologies in their field, it is essential to present a toolkit that fulfils their requirements to freely access the data store without the need to worry about its underlying technology.

In this demo, we introduce ForestQB, a tool that aims to facilitate the knowledge extraction out of the RDF triplestores by allowing the researchers to construct their query visually. The tool provides a high level of abstraction for users by supporting a Form-based interface with an integrated conversational AI to support natural query construction.

\section{ForestQB Features}
ForestQB is a web application that works on the browser to query and explore RDF triplestores that were exposed as a SPARQL endpoint. ForestQB was implemented based on the requirements collected by reviewing the current progress in the field and interviewing experts in the Bioscience field (the stakeholder). The recent features of the tool can be outlined as follows:

\begin{itemize}
  \item \textbf{General overview}: the initial interface provides a basic search functionality by allowing the user to select sensors (observable properties) as a starting point to explore the data. The interface contains a list of sensors loaded from the endpoint when the toolkit was initially loaded. The sensors' data are fetched by sending a SPARQL query that can be adjusted from the settings to customise the tool if querying different endpoints. Hence, The user can click the search button to start loading the results related to that sensor without requiring additional steps. In addition, the interface includes a map with geospatial filtering capabilities by drawing circles directly on the map. It also contains an upper date picker that will strict the results to a general date range. The upper left side includes a list of predefined examples that will populate all required fields to promote learning by example. The primary goal is to have a user-friendly interface that will support the user to explore the data quickly.
  
  \item \textbf{Detailed querying}: once the users decide to have more advanced querying, they can click on the search customisation to display all the detailed querying features. The interface will display all the linked properties for the chosen sensor with an additional sublist allowing them to manually add filters to each property. The filters that the user can add will vary based on the property XSD data type. For example, a string will include Contain, Match and Regex filters. Also, the same filters will act differently based on the property data type. For example, the number and date both have a Range filter, but while the number will show a numeric text field, the date will show a date picker. Furthermore, the user can choose to hide the entity from the results or mark an entity as optional to be ignored if it does not exist.
  
  \item \textbf{Conversational AI}: the ForestQB includes a component that will display a chatbox where the user can express their query in natural language to populate the corresponding fields automatically. In addition to constructing queries, it can help the user to inquire about the underlying data structure. For example, the chatbot can answer simple questions such as "What are the sensors?", "What is Aqeela?" and "Where is Aqeela?" as shown in Figure \ref{fig:figure2}. The user can completely hide the form-based interface to use the chatbot as a stand-alone query builder tool to query and retrieve the results.

\end{itemize}

\section{Our SPARQL endpoint}
Our dataset contains sensitive historical data collected by bioscientists that were modelled as Linked-data. Thus, our SPARQL endpoint is privately available. The data integrates multiple ontologies to define its underlying structure, including SOSA \cite{Janowicz2019} ontology. Hence, ForestQB initially relies on SOSA to get all connected sensors and their observable properties, which means it will potentially work with another endpoint if it applies SOSA.
\begin{figure}[t]
\includegraphics[scale=0.30]{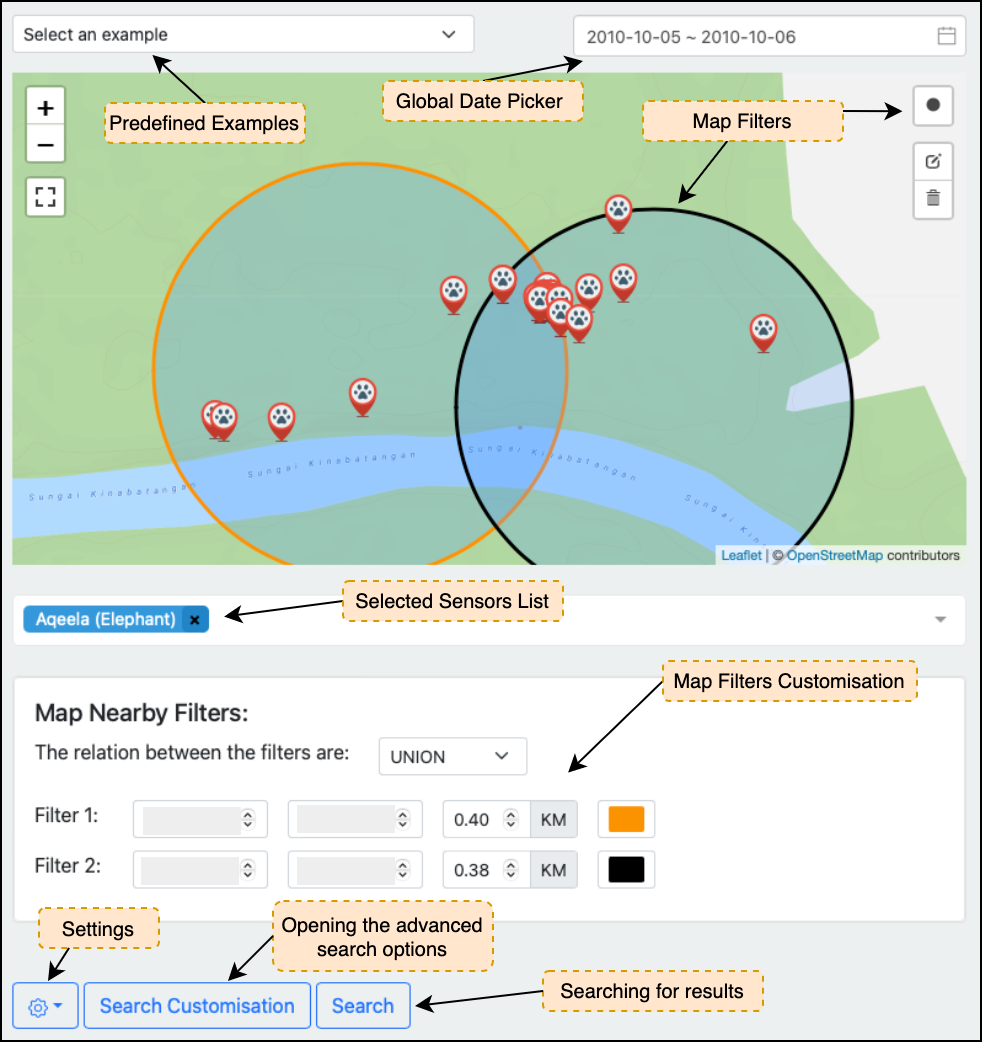}
\centering
\caption{Overview of a simple query to retrieve data of a particular animal within a specific date and location. The Longitude and Latitude of the animals were hidden due to the sensitivity of the data.}
\label{fig:figure1}
\end{figure}

\section{Demonstration}
The demo will illustrate the ForestQB interface and explain its functionality. Figure \ref{fig:figure1} shows an overview of the tool to create a simple query using map filters, demonstrating the basic search with a limited number of filters to apply. The \quotes{customisation search} button is where the user can apply a more advanced search. The customisation is always hidden until the user decides to display it to allow a more neat look and hide the complexity of the interface. The results will be presented in a tabular format to reflect the user query.

\begin{figure}[t]
\includegraphics[scale=0.3]{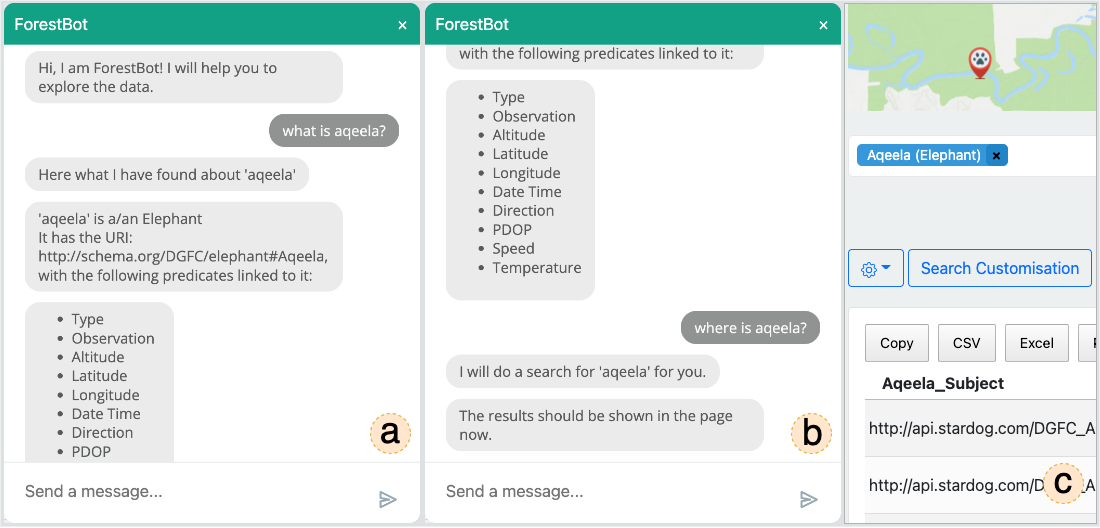}
\centering
\caption{Overview of the conversational AI. (a) Retrieving information about \quotes{Aqeela} sensor. (b) Constructing a query to find \quotes{Aqeela}. (c) The result will then be reflected on the Query Builder as if the user has selected them, and the search process will be triggered.}
\label{fig:figure2}
\end{figure}

The conversational AI can work as an assistive tool to the Form-based builder or as a standalone query builder. The user can type their query in natural language, and it will be reflected on the interface. For example, a question such as \quotes{Where is aqeela?} will select the correct sensor from the sensors list and trigger the search process (see Figure \ref{fig:figure2}).

\section{Technical Details}
The ForestQB interface was designed after reviewing the current state-of-the-art and gathering requirements from stakeholders. The design has numerous layers and components, each of which serves a unique purpose with distinct technical details behind it. Thus, this section briefly discusses some of the technical aspects of ForestQB.
\begin{list}{\labelitemi}{\leftmargin=1em}
%   \item \textbf{Query Builder}:
%   \item \textbf{Conversational AI}:
%   \item \textbf{Converting UI choices onto SPARQL}:

\item \textbf{Web Application}:
Plain JavaScript and Vue.js\footnote{https://vuejs.org/} Framework have been used to build the tool as a Single-page application. The tool will generate a JSON object that will be identical to the tool choices. Once the user clicks search, this object will be sent to the conversion engine (as an AJAX request) to be converted to SPARQL and sent back to the tool to be used for querying the endpoint. The separation of the SPARQL conversion mechanism from the tool will allow it to be shared between the ForestQB and the conversational AI. Both ForestQB and conversational AI share a centralised store using Vuex\footnote{https://vuex.vuejs.org/}.

\item \textbf{Conversational AI model}:
The conversational AI is split into two pieces: the front-end and the classification model. The front-end is part of the main web application components. However, the Natural Language Understanding (NLU) model has been built using RASA\footnote{https://rasa.com/} framework. Thus, all of the logic behind the NLU lies under a different web server that is powered by RASA. The front-end will send all user messages to the chatbot server to understand the user's intent. Most of the actions (responses) are implemented within the web front-end. The conversational AI will adjust the centralised store based on the classified intent and entities to modify the JSON Query object. Then, it will trigger the search process. The chatbot will either answer simple questions about the data schema or retrieve the results when the user asks.

\item \textbf{Map Filters}:
ForestQB is offering map filters to narrow down the results. The implemented map uses the Leaflet\footnote{https://leafletjs.com/} library to visualise and allow the map's main functionality. Thus, drawing circles on the map allows the user to create nearby filters. Once a circle is drawn, its coordinates and parameters are saved on the JSON Query object and then translated into SPARQL. In addition, the user can define the relationship between the circles as intersections or unions.
\end{list}

\section{Conclusions}
This paper has briefly introduced the ForestQB, a toolkit that aims to assist bioscientists in querying Linked-Data. The tool supports generating the query by mixing the conversational AI with the Form-Based query builder. The ForestQB is still a work in progress, as we plan to improve its conversational AI to handle more complex sentences. In addition, it will support more visualisation options based on our future user study.

%
% ---- Bibliography ----
%
% BibTeX users should specify bibliography style 'splncs04'.
% References will then be sorted and formatted in the correct style.
%
\bibliographystyle{splncs04}
\bibliography{refs}
%
% \begin{thebibliography}{8}

% \bibitem{ref_article1}
% Author, F.: Article title. Journal \textbf{2}(5), 99--110 (2016)

% \bibitem{ref_lncs1}
% Author, F., Author, S.: Title of a proceedings paper. In: Editor,
% F., Editor, S. (eds.) CONFERENCE 2016, LNCS, vol. 9999, pp. 1--13.
% Springer, Heidelberg (2016). \doi{10.10007/1234567890}

% \bibitem{ref_book1}
% Author, F., Author, S., Author, T.: Book title. 2nd edn. Publisher,
% Location (1999)

% \bibitem{ref_proc1}
% Author, A.-B.: Contribution title. In: 9th International Proceedings
% on Proceedings, pp. 1--2. Publisher, Location (2010)

% \bibitem{ref_url1}
% LNCS Homepage, \url{http://www.springer.com/lncs}. Last accessed 4
% Oct 2017
% \end{thebibliography}
\end{document}